# Giant thermoelectric response of confined electrolytes with thermally activated charge carrier generation

Rajkumar Sarma and Steffen Hardt[*]

Technische Universität Darmstadt, Fachbereich Maschinenbau, Fachgebiet Nano- und Mikrofluidik

[*]hardt@nmf.tu-darmstadt.de

**Abstract**: The thermoelectric response of thermally activated electrolytes (TAE) in a slit channel is studied theoretically and by numerical simulations. The term TAE refers to electrolytes whose charge carrier concentration is a function of temperature, as recently suggested for ionic liquids and highly concentrated aqueous electrolyte solutions. Two competing mechanisms driving charge transport by temperature gradients are identified. For suitable values of the activation energy that governs the generation of charge carriers, a giant thermoelectric response is found, which could help explaining recent experimental results for nanoporous media infiltrated with TAEs.

In the past few years, a number of materials were reported that show a very pronounced thermoelectric response. The thermoelectric response of a material is usually quantified in terms of the Seebeck coefficient $S = \Delta V/\Delta T$, which measures the thermovoltage generated per temperature difference applied across a layer of material. Remarkably, giant Seebeck coefficients were measured mainly in materials in which an electrolyte fills a confined space, for example the pore space of a nanoporous material. Examples are carbon nanotube-based supercapacitor materials [1] or cellulose-based membranes [2,3] infiltrated with aqueous electrolyte solutions, and polymer matrices infiltrated with ionic liquids [4–6]. In units of $k_B/e$ ($k_B$: Boltzmann constant, $e$: elementary charge), the Seebeck coefficients measured in such systems may reach values of more than 500. It was pointed out in [7] that such high values stand in contradiction to the classical theory of thermodiffusion in dilute electrolyte solutions, which predicts Seebeck coefficients of the order of $k_B/e$. Confinement effects help increasing the Seebeck coefficient beyond this level even when applying the standard Poisson-Nernst-Planck (PNP) theory for dilute electrolytes [8]. However, the giant thermovoltages measured in the complex electrolyte systems referred to above cannot be explained based on the standard PNP equations.

To explain the physics behind the huge Seebeck coefficients, in [7] a charge transport mechanism was suggested that is fundamentally different from that in dilute electrolyte solutions. According to that, the charge separation occurs by thermally-induced hopping of charge carriers between the minima of a periodic potential energy landscape. In the present work, we suggest an alternative mechanism behind the huge Seebeck coefficients.

The key concept this description hinges on is thermally activated charge carrier generation. In the past few years, it was observed that especially in confined ionic liquids, the Debye length is much larger than what the ion concentration appears to suggest [9,10]. The same seems to apply to highly concentrated aqueous electrolyte solutions [11]. This phenomenon has been referred to as "underscreening", and a number of attempts have been made to elucidate the physics behind it. Underscreening also occurs in unconfined ionic liquids [11], but confinement appears to introduce some additional effects that reduce the concentration of effective charge carriers (ECCs). A possible explanation for underscreening is that ions can either be found in a free or a bound state, the latter being a cluster of strongly correlated ions [12,13]. Consequently, an ion bound to such a cluster no longer acts as an ECC, but large charge-neutral clusters containing many ions may occur. To describe the equilibrium between ions in clusters and those acting as effective charge carriers, a Boltzmann distribution is employed [10–12], which implies that the concentration of ECCs has a strong temperature dependence. This temperature dependence is the cornerstone of the model we have developed for thermoelectricity in confined ionic liquids and concentrated aqueous electrolyte solutions. For brevity, in the following we will refer to these classes of electrolytes as "thermally activated electrolytes" (TAE).

Figure 1a schematically illustrates the system under consideration. A narrow confinement of half-width $h$ and length $l\left(h/l = A\right)$ is filled with a TAE. The ECCs form an electric double layer (EDL) at the charged channel walls. It is the axial concentration gradient depicted in Figure 1b that drives charge transport through the channel and is responsible for the thermoelectric effect. It can be seen that the magnitude of the concentration gradient in a TAE (yellow solid line) is larger than in a scenario in which the charge carrier concentration is constant (green dashed line), as further discussed below. Only the latter is described by the standard PNP equations, as analyzed in previous work [8]. In addition, in a TAE the sign of the concentration gradient is usually reversed compared to the standard scenario.

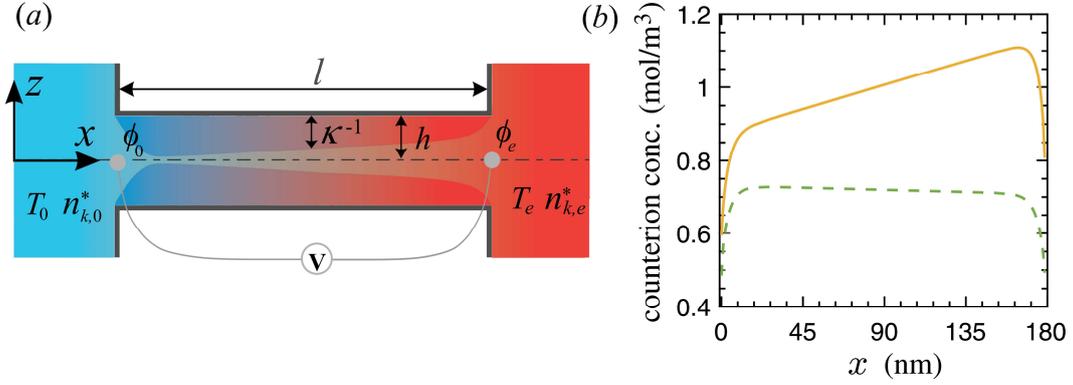

FIG. 1. (a) Schematic of a narrow-confined space of half-width $h$, bounded by two parallel plates and connecting two reservoirs maintained at temperatures $T_0$ and $T_e (= T_0 + \Delta T)$, with subscripts $0$ and $e$ denoting the properties at the left (cold) and the right (hot) reservoir, respectively. Debye layers with a characteristic temperature-dependent thickness of $\kappa^{-1}$ form at the channel walls. (b) An axial gradient in the cross-section averaged counterion concentration develops through the superposition of two competing effects. First, there is the temperature dependence of the wall-normal diffusion and electromigration fluxes, which are related to each other by the Einstein-Smoluchowski equation. The temperature dependence already exists at constant effective charge carrier concentration (green dashed line). This effect, however, is overridden when in addition the temperature-dependent effective charge carrier concentration (yellow solid line) is considered. The $x$-axis extends over the entire length of the channel.

To model the transport processes in the channel, we assume a small enough temperature gradient, such that we can apply the framework of classical irreversible thermodynamics [14]. The key assumption in this context is that at each position inside the channel, the state of the system can be well approximated based on local thermodynamic equilibrium. We consider a three-component system, a background fluid together with two types of ECCs, positive and negative ones, and suppose that the molar concentration of the ECCs is significantly smaller than that of the background fluid. Thereby we aim at providing a model for a TAE in which ECCs are created in a thermally activated process [10–12]. For example, in the case of an ionic liquid, the background fluid would consist of charge-neutral clusters of strongly correlated ions from which charge carriers emerge at sufficiently high temperatures. In such a scenario, there could be multiple types of anionic and cationic ECCs, but to keep our model simple and in the spirit of a proof-of-principle study, we limit ourselves to one type of anionic and one type of cationic ECCs.

Under these conditions, the standard description of species transport relies on the Nernst-Planck equations (NPE), comprising diffusive, advective and electromigration contributions in the ECC flux density $\mathbf{j}_k$, with $k = (+,-)$. In the stationary state the NPE reduce to $\nabla \cdot \mathbf{j}_k = 0$. The NPE are supplemented by the Navier-Stokes equations (for the velocity field in the fluid), the energy equation (for the temperature field), and the Poisson equation (for the electric potential). Details concerning the set of underlying equations and the corresponding boundary conditions can be found in the supplemental material [15]. Note that the underlying NPE differs from the conventional version found in the literature by the fact that the charge carrier (ion) concentration is not given as a boundary condition, but adjusts itself according to the local temperature. Effectively, the model comprises a chemical equilibrium between bound and free charges that is attained instantaneously. The mathematical analysis is based on the leading-order contribution of the governing equations under the thin-gap approximation $(A \ll 1)$, signifying that the dimensionless axial gradient of any variable is much smaller than its transverse counterpart. The energy equation forms the only exception in that context, i.e. the thin-gap approximation is not applied to this equation. Under this approximation, it is shown in the supplemental material ([15], with [16–26] cited therein) that the effects of advection, viscous dissipation and Joule heating can be neglected in the transport equations. Hence, the Nernst-Planck flux becomes

$$-\mathbf{j}_k = D_k \nabla n_k^* + e v_k n_k^* \omega_k \nabla \phi, \qquad (1)$$

where for the valence of the ECCs $v_\pm = \pm v$ is assumed. The molar concentration of the ECCs is given by $n_k^*$. The Einstein-Smoluchowski equation $\omega_k = D_k / k_B T$ relates the electrophoretic mobility ($\omega_k$) of the ECCs to their diffusion coefficient ($D_k$). Thermodiffusion effects are neglected, owing to their usually small contribution to the induced thermoelectric field in a highly confined space [8]. $\phi = \psi + \phi_{in}$ is the total electrostatic potential, the sum of the EDL potential $\psi$ and the induced thermoelectric potential $\phi_{in}$. The electrostatic potential is governed by the Poisson equation

$$\nabla \cdot (\varepsilon \nabla \phi) = -\rho_f, \qquad (2)$$

where $\varepsilon$ is the dielectric permittivity of the fluid and $\rho_f = \sum ev_k n_k^*$ is the volumetric charge density.

In terms of the leading-order contribution in the small parameter $A$, the temperature gradient in the fluid is given by $\nabla T = (\Delta T/l, 0)$ ([15], Sec.1). Based on that and the Boltzmann distribution determining the concentration of ECCs [10–12], their concentration will vary not only due to the electrostatic potential distribution, but also due to the local temperature. This is the essential mechanism driving the thermoelectric effect. Omitting terms $\mathcal{O}(A^2)$ in the NPE, the ECC distribution follows from $j_{k,z} = 0$. The result is ([15], Sec.2)

$$n_k^* = n_b \exp\left(-\frac{E_d + ev_k \psi}{k_B T}\right). \qquad (3)$$

The bulk concentration far away from solid walls is given as $n_k^* = n_b \exp(-E_d/k_B T)$. $E_d$ is the activation energy needed to create ECCs from ion clusters [10–12]. In the limit $E_d \to 0$, Eq. (3) reduces to the well-known Boltzmann distribution valid for an isothermal channel. Inserting (3) in (2) and keeping only leading-order terms in $A$ ([15], Sec. 3) yields the Poisson-Boltzmann equation, i.e. $d^2\Psi/dz^2 = \kappa^2 \sinh(\Psi)$, where $\Psi = ev\psi/k_B T$ is the non-dimensional electrostatic potential. The Debye parameter $\kappa = \sqrt{2e^2 v^2 n_b \exp(-E_d/k_B T)/\varepsilon k_B T}$ becomes a local quantity, depending on $T(x)$. For $\Delta T/T_0 \ll 1$, $\kappa$ can be rewritten as ([15], Sec. 3)

$$\kappa = \kappa_0 + \frac{1}{2}\frac{E_d(T-T_0)}{k_B T_0^2}\kappa_0 - \frac{1}{2}\frac{(T-T_0)}{T_0}\kappa_0. \qquad (4)$$

The first term on the right-hand-side of (4) denotes the Debye parameter at the channel entrance, the second term accounts for the variation of $\kappa$ caused by thermally-induced charge carrier generation, while the last one depicts the variation caused by the temperature dependence of $D_k/\omega_k$. The opposite signs of the second and third term in (4) signify their counteracting contributions with respect to the EDL thickness. The corresponding electrostatic potential can be obtained by solving the Poisson-Boltzmann equation in the Debye-Hückel (DH) limit $(|\Psi|<1)$, using the boundary conditions $\psi = \zeta$ at $z = \pm h$. The result is $\psi_{DH} = \zeta \cosh(\kappa z)/\cosh(\kappa h)$. Combining (1) and (3), the axial ECC flux densities are obtained as

$$-\frac{j_{k,x}}{n_k D_k} = \left(\frac{E_d}{k_B T^2} + \frac{e v_k \psi}{k_B T^2}\right)\frac{dT}{dx} - \frac{e v_k}{k_B T} E, \quad (5)$$

where $E = -\partial \phi_{in}/\partial x$. The induced electric field $E$ can be computed by equating the total electric current $I = \int_0^h \sum_{k=1}^N e v_k j_{k,x} dz$ to zero. Assuming $D_k = D$, the Seebeck coefficient $S = E/(dT/dx)$ reads ([15], Sec. 4)

$$S = -\frac{E_d}{evT}\frac{\int_0^h \sinh\left(\frac{ev\psi}{k_B T}\right)dz}{\int_0^h \cosh\left(\frac{ev\psi}{k_B T}\right)dz} + \frac{1}{T}\frac{\int_0^h \psi \cosh\left(\frac{ev\psi}{k_B T}\right)dz}{\int_0^h \cosh\left(\frac{ev\psi}{k_B T}\right)dz}. \quad (6)$$

Evaluation of the integrals under the DH approximation yields

$$S_{DH} = \frac{-\dfrac{E_d \zeta}{k_B T^2}\dfrac{\tanh(\kappa h)}{\kappa h} + \dfrac{\zeta}{T}\dfrac{\tanh(\kappa h)}{\kappa h}\left[1 + \dfrac{1}{2}\left(\dfrac{ev\zeta}{k_B T}\right)^2\left\{\dfrac{\tanh^2(\kappa h)}{3} + \dfrac{1}{\cosh^2(\kappa h)}\right\}\right]}{1 + \left(\dfrac{ev\zeta}{2k_B T}\right)^2\left\{\dfrac{\tanh(\kappa h)}{\kappa h} + \dfrac{1}{\cosh^2(\kappa h)}\right\}}. \quad (7)$$

The expression for $S_{DH}$ remains accurate only for $|ev\zeta/k_B T| = |\bar{\zeta}| < 1$. For small $\zeta$ and $\Delta T$ (such that $|\bar{\zeta}| < 1$ and $\Delta T/T_0 \ll 1$), $S_{DH}$ can be further simplified to give ([15], Sec. 4)

$$S_{DH}\frac{T_0}{\zeta} = -\left(1 - 2\frac{\Delta T}{T_0}\right)\frac{E_d}{k_B T_0}\frac{\tanh(\kappa h)}{\kappa h} + \left(1 - \frac{\Delta T}{T_0}\right)\frac{\tanh(\kappa h)}{\kappa h}. \quad (8)$$

Equations (6)-(8) depict that the net thermoelectric field for confined TAEs is due to the superposition of two different effects. One is an axial gradient of $n_k^*$ affiliated with the temperature-modulated dissociation of ionic aggregates; the other one is the gradient in $n_k^*$ resulting from the temperature dependence of $D_k/\omega_k$. Specifically, the Seebeck coefficient of Equation (8) consists of two different contributions. The second term on the right-hand side is the only contribution obtained for a dilute electrolyte (DE), as considered in [8]. Here, the term DE refers to an electrolyte in the conventional sense, i.e. with an ion concentration defined by the boundary conditions and small compared to the solvent concentration. By contrast, in a TAE, the charge carrier concentration is not given but results from the dissociation of charge-neutral ionic aggregates. For TAEs, the first term on the right-hand side of Equation (8) is present in addition to the second one. As we will show in the following, the first term will usually dominate the thermoelectric response.

To apply our model to a specific system, we focus on the TAE ([C$_2$mim][NTf$_2$]) considered in [10], which is an ionic liquid. To the best of our knowledge, the transport coefficients of the ECCs in this medium are unknown. However, the Seebeck coefficient only depends on the ratio $D_k/\omega_k = k_B T$, i.e. the results are independent of the specific values of the transport coefficients. In addition to the diffuse part of the EDL, in [10] the existence of a Stern layer was reported, which is not included in our model description. In the supplemental material ([15], Sec. 5), we estimate the effects due to the Stern layer. The results indicate that its presence reduces the Seebeck coefficient. However, the transport processes inside the Stern layer most probably lie beyond the realm of continuum theory [27]. Moreover, the atomistic mechanism suggested in [7], i.e. thermally-induced hopping of charge carriers, could result in an increase of the Seebeck coefficient due to the Stern layer if the solid surface provides a suitable potential energy landscape. This leaves us with the conclusion that currently, no clear statements about the effects due to the Stern layer are possible, and in the spirit of a proof-of-principle study, we focus on charge transport inside the diffuse part of the EDL. For the conditions considered below, the thickness of the diffuse part lies between 6 and 7 nm [10].

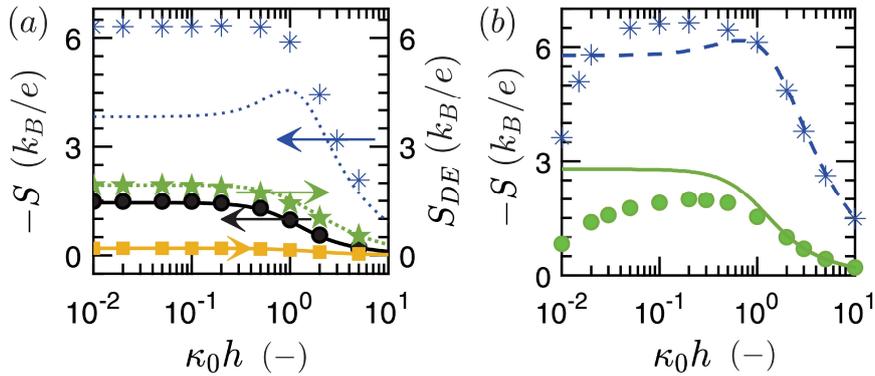

FIG. 2. (a) Seebeck coefficient (in units of $k_B/e$) as a function of the non-dimensional Debye parameter $(\kappa_0 h)$ for a TAE $(-S)$ and a DE $(S_{DE})$. Results based on the DH approximation (lines) are compared with results from the full numerical simulation of the coupled PNP equations (symbols). The solid lines (and the corresponding symbols) correspond to $\zeta = 5$ mV, the dotted lines (and the corresponding symbols) represent $\zeta = 50$ mV. (b) Dependence of the Seebeck coefficient on the wall boundary conditions. A constant surface charge density $q$ (symbols) or $\zeta$ potential (lines) is imposed at the bounding walls. Green circles: $q = -0.27 \times 10^{-4}$ C/m$^2$, Blue stars: $q = -2.5 \times 10^{-4}$ C/m$^2$; Green solid line: $\zeta = 10$ mV, Blue

dashed line: $\zeta = 70\,\text{mV}$. The activation energy considered here corresponds to [C$_2$mim][NTf$_2$] [10].

Figure 2a compares the Seebeck coefficient of TAEs $(-S)$ and DEs $(S_{DE})$, given in units of $k_B/e$, as a function of $\kappa_0 h$ for $\zeta = \{5, 50\}\,\text{mV}$. Unless specified otherwise, we consider $T_0 = 298\,\text{K}$, $\Delta T = 25\,\text{K}$, $E_d = 9.4 k_B T$, $\varepsilon = 12.3$ and $l = 250 h$. These particular values of $E_d$ and $\varepsilon$ correspond to the TAE [C$_2$mim][NTf$_2$] [10]. In the case of a DE, we chose $\varepsilon = 78.5$, corresponding to the permittivity of water at a temperature of 298 K. Our analytical results are juxtaposed with those based on a numerical solution of the coupled PNP equations ([15], Sec. 8), obtained using the finite element simulations. For DEs, any difference between the two classes of results is practically indistinguishable. By contrast, for TAEs, while the PNP solutions agree well with the DH approximation at small $\zeta\,(=5\,\text{mV})$, the latter significantly underestimates $S$ at higher values of $\zeta$. The exclusion of $\mathcal{O}(\psi^n, n \geq 3)$ terms in the integrand of (6) is responsible for this deviation.

One can readily observe that irrespective of the value of $\kappa_0 h$, the TAE yields a significantly higher Seebeck coefficient (in terms of its magnitude) compared to the DEs. The opposite sign of $S$ between the TAE and the DEs indicates opposite directions of the induced thermoelectric field. For $E_d \to 0$ and $\Delta T > 0$, Eq. (4) shows that the EDL expands in the direction of increasing temperature, implying a reduced counterion concentration. On the other hand, for TAEs, the concentration gradient of ECCs due to the applied temperature gradient keeps reducing the EDL thickness in the direction of increasing temperature. This can be verified from Eq. (4) for $E_d > 0$. Although $D_k/\omega_k$ is a function of temperature for TAEs as well, the temperature-driven ECC generation dominates and establishes a net concentration gradient opposite to that caused by the temperature-dependent $D_k/\omega_k$. The magnitude of $S$ is at its maximum for $\kappa_0 h \to 0$, i.e. in the regime of highly overlapping EDLs, in agreement with the results reported in [8]. In this case, we obtain $S_{DH} = \zeta(1 - 2E_d/\mathcal{R})/T$ with $\mathcal{R} = k_B T (2 + \bar{\zeta}^2)$, based on the DH approximation. By contrast, $S_{DH} \to 0$ for $\kappa_0 h \to \infty$.

To check the robustness of our results, it is helpful to check their sensitivity with respect to the underlying model assumptions. One important assumption is the Dirichlet boundary condition for the electric potential at the channel walls. Considering the plethora of combinations of solids and TAEs, it is unclear how viable this assumption is. A more realistic boundary condition could be obtained from a charge regulation model, which describes the chemistry of charge formation at the channel walls. However, formulating a charge regulation model that is valid for a broad range of material combinations is a close-to-impossible task. In [28], the constant potential boundary condition was compared with the constant surface charge density and a charge regulation boundary condition. The main results were: (i) For $\kappa_0 h \gg 1$, the results become independent of the specific boundary condition; (ii) The data obtained with the most realistic model (charge regulation) are always bracketed by the data for constant potential and constant wall charge density. Motivated by these results, we have repeated our calculations for constant wall charge density $q$. The mapping between $\zeta$ and $q$ is achieved following [15], Sec. 6 ([29,30]). The corresponding results were numerically computed by solving the PNP equations ([15], Sec. 8). For a constant $\zeta$ potential, $S$ was numerically obtained by solving Eq. (6).

The resulting Seebeck coefficient is displayed in Figure 2b as a function of $\kappa_0 h$. In the regime of non-overlapping EDLs $(\kappa_0 h \gg 1)$, $S$ remains unaffected by the specific type of boundary condition. However, for overlapping EDLs, the results based on these two different boundary conditions start deviating, with increasing deviations as $\kappa_0 h \to 0$. This disagreement can be explained by the fact that for $A \ll 1$, the charges in the liquid are exactly equal in magnitude to the wall charges, ruling out any axial concentration gradients as $\kappa_0 h \to 0$. Importantly, for $\kappa_0 h \approx 1$, the difference between the two sets of results is rather small. Considering the typical functional dependence of $S$ with $\kappa_0 h$, we conclude that at large $\zeta$ potentials, the Seebeck coefficient evaluated at $\kappa_0 h = 1$ is a reasonably good indicator for the maximum Seebeck coefficient that can be achieved (independent of the wall boundary condition).

So far, we have only considered one specific TAE. Owing to the numerous types of ionic liquids [31] and concentrated electrolytes, one can expect a broad range of activation energies $E_d$. Figure 3a shows the Seebeck coefficient, $-S$ (in units of $k_B/e$) as a function of $E_d$ (in units of $k_B T$) for different $\zeta$ potentials. The values considered here for the parameters $E_d$ and $\zeta$ are in accordance with the experimental data reported for TAEs [10,32]. Results obtained using the DH approximation are compared with the numerical evaluation of Eq. (6) ([15], Sec. 7). Both $\zeta$ and $E_d$ have a substantial influence on $S$, with a sign reversal of $S$ occurring within $1.1 \leq E_d/(k_B T) \leq 4$ for the range of $\zeta$ potentials considered. Defining $dn_k^*/dT$ as the temperature sensitivity parameter, it is found that small $E_d$ causes low temperature sensitivity despite a high ECC concentration. At small $E_d$, therefore, the thermoelectric field is governed by the concentration gradient due to the temperature-dependent $D_k/\omega_k$, yielding a Seebeck coefficient which differs qualitatively from that for higher $E_d$. On the other hand, with increasing $E_d$, the concentration gradient is more and more dominated by the temperature-dependent generation of ECCs, and eventually results in a Seebeck coefficient that is much higher compared to that in DEs. Despite evidence that $E_d$ can attain values as high as $100 k_B T$ [32], in Figure 3a we have limited the range of activation energies to $40 k_B T$. The underlying reason is that for high activation energies, the charge density drops to values that are probably physically insignificant. $E_d \approx 40 k_B T$ marks the limit that corresponds to the charge density in liquids with very low conductivity [33,34], which we consider as a physically reasonable point of reference. Figure 3b shows the variation of the cross-section averaged counterion concentration $(\bar{n}_c)$ and electric potential $(\bar{\psi})$ along the channel for vanishing voltage between the reservoirs and different $E_d$, as obtained from numerical calculations. It is apparent that $\bar{n}_c$ drops appreciably with increasing $E_d$. Apart from the counterion concentration gradient, also the intrinsic axial electric field (represented by the negative gradient of $\bar{\psi}$), induced by the temperature gradient, assists in driving the counterion flux. However, this effect is minor compared to counterion diffusion in the concentration gradient. The thermovoltage due to the counterion flux depends on an interplay between the average counterion concentration (decreasing the thermovoltage by increasing the conductivity inside the channel) and the concentration gradient (increasing the thermovoltage). Within the considered range of $E_d$ values, the Seebeck coefficient increases with $E_d$ and reaches very large values. The underlying mechanisms could help explaining experimental data obtained with highly concentrated electrolytes or ILs in nanoconfinement [3,4,35,36] that point towards Seebeck coefficients much in excess of $k_B/e$.

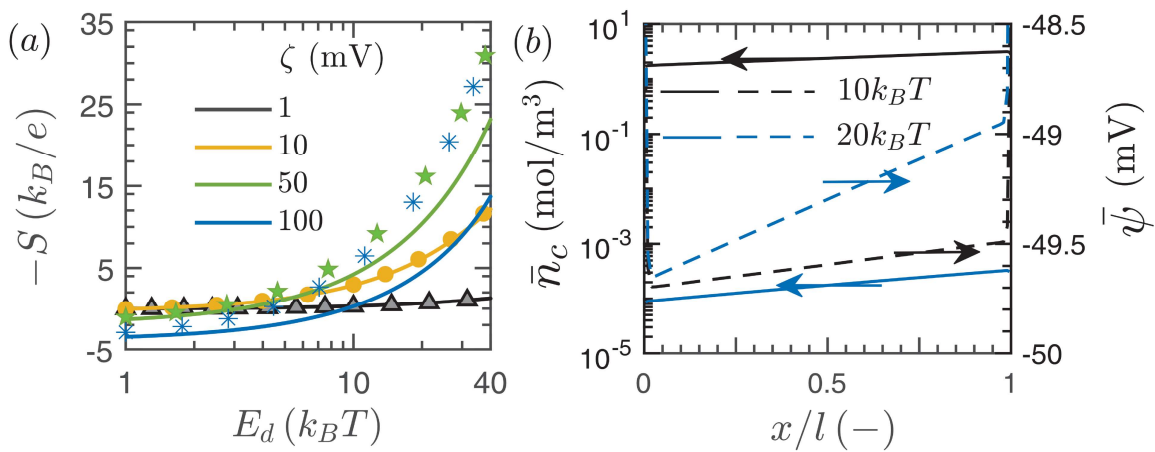

FIG. 3. (a) Dependence of the Seebeck coefficient on the activation energy and $\zeta$ potential at $\kappa_0 h = 0.1$. Results based on the numerical evaluation of Eq. (6) (symbols) are compared with the DH approximation (solid lines). (b) Axial variation of the cross-section averaged counter-ion concentration and electric potential, as obtained from numerical calculations. Solid lines represent $\bar{n}_c$, while the dashed lines represent $\bar{\psi}$. No voltage was applied between the reservoirs, and $\kappa_0 h = 0.1$, $\zeta = -50$ mV were chosen.

In conclusion, we have shown that the Seebeck coefficient in confined electrolytes can reach values much larger than $k_B/e$. Such a giant thermoelectric response is found for thermally activated electrolytes, in which charge carriers are generated in a thermally activated process. The temperature dependence of the charge carrier concentration is usually the dominant mechanism driving charge transport. This mechanism can help explaining the giant Seebeck coefficients that were measured in materials in which an electrolyte fills a confined space.

We consider our work a proof-of-principle study based on a comparatively simple model that includes a single species of effective charge carriers. It cannot be expected that this model captures all transport process occurring in ionic liquids or concentrated aqueous electrolytes. Examples for additional effects that could be the subject of future studies are the thermally induced hopping of charge carriers [7], ion transport inside the Stern layer, the intrinsic thermophoretic mobility of the effective charge carriers, multicomponent diffusion in the case of multiple species of charge carriers, or conversion of different types of charge carries into each other. Some of these phenomena could lead to further enhancements of the thermoelectric response and thereby contribute to understanding the giant response of some porous, electrolyte-filled materials.

**Funding acknowledgement**

This project has received funding from the European Union's Horizon 2020 research and innovation program under grant agreement No 964251.

# Giant thermoelectric response of confined electrolytes with thermally activated charge carrier generation: Supplemental Material

The theoretical modeling of the transport processes associated with thermally activated electrolytes (TAE) involves the Nernst-Planck equations (NPE), the Navier-Stokes equation (NSE), the energy equation and the Poisson equation. The NPE describes the evolution of the charge carrier concentration field $n_k^*$ and is given by

$$\frac{\partial n_k^*}{\partial t} + \mathbf{u} \cdot \nabla n_k^* + \nabla \cdot \mathbf{j}_k = \mathcal{G}_k. \tag{S1}$$

In (S1), $\mathbf{u}$ is the flow velocity vector, $\mathbf{j}_k$ is the ion flux density defined by (1), while $\mathcal{G}_k$ is the rate of generation of effective charge carriers (ECCs) of species $k$ due to the dissociation of ionic aggregates. For an incompressible electrolyte, the NSE takes the following form

$$\rho \left[ \frac{\partial \mathbf{u}}{\partial t} + (\mathbf{u} \cdot \nabla) \mathbf{u} \right] = \nabla \cdot \mathcal{T}_V + \nabla \cdot \mathcal{T}_M \tag{S2}$$

where $\rho$ is the mass density. In (S2), $\mathcal{T}_V = -p\mathbf{I} + \tau_{ij}$ represents the mechanical stress tensor, the sum of the pressure term $p\mathbf{I}$ ($\mathbf{I}$ denotes the unit tensor) and the viscous stress $\tau_{ij} = \eta \left[ \nabla \mathbf{u} + (\nabla \mathbf{u})^T \right]$, where $\eta$ is the dynamic viscosity. Furthermore, $\mathcal{T}_M = \varepsilon \left( \nabla \phi \nabla \phi - \nabla \phi \cdot \nabla \phi \mathbf{I}/2 \right)$ denotes the Maxwell stress tensor [S1], with $\phi$ as the electric potential and $\varepsilon$ as the dielectric permittivity. $\phi$ is determined by the Poisson equation

$$\nabla \cdot (\varepsilon \nabla \phi) = -\rho_f. \tag{S3}$$

In (S3), $\rho_f = \sum_{k=1}^{K} e v_k n_k^*$ is the volumetric charge density of the ECCs. The energy equation reads

$$\rho c_p \left[ \frac{\partial T}{\partial t} + \mathbf{u} \cdot \nabla T \right] = \nabla (\mathcal{K} \nabla T) + \dot{\mathcal{Q}}_\phi + \dot{\mathcal{Q}}_\eta, \tag{S4}$$

where $c_p$ is the specific heat capacity (at constant pressure) and $\mathcal{K}$ is the thermal conductivity. The second and third terms on the right-hand side (RHS) of (S4) are the contributions of Joule heating and viscous dissipation, respectively.

To obtain a closed-form expression for the Seebeck coefficient $S$, the mathematical model needs to be supplied with boundary conditions. For the NPE, the wall boundaries (located at $z = \pm h$) are considered to be mass impermeable, $j_{k,z} = 0$, while a fixed $\zeta$ potential is specified at the walls for analytically solving the Poisson equation. Alternatively, a fixed surface charge density $q$ at the channel walls is considered, essentially to compare the Seebeck coefficients obtained under a specified $\zeta$ potential and a specified charge density $q$. For the temperature field, thermally insulated walls ($\partial T/\partial z\big|_{z=\pm h} = 0$) are considered, while for the velocity field, the no-slip boundary condition $\left(u\big|_{z=\pm h} = 0\right)$ is applied at the walls.

## §1: Temperature distribution within the confinement

To compute the temperature distribution within the confinement, first one needs to assess the influence of Joule heating and viscous dissipation on the temperature field. The Joule heating term can be approximated as $\dot{Q}_\phi = \sigma_B E^2$, where $\sigma_B \left(= 2e^2 v^2 n^* D/k_B T\right)$ is the bulk electric conductivity. For fully developed flow through a slit channel, $\mathbf{u}$ becomes $\mathbf{u} = (u,0)$. The heat generation through viscous stresses is, therefore, quantified by $\dot{Q}_\eta = \eta \left(\partial u/\partial z\right)^2$. Introducing the non-dimensional variables $(X,Z) = (x/l, z/h)$, $\bar{u} = u/U$, $\bar{t} = tU/l$ and $\theta = (T - T_r)/\Delta T$, where $U$ is a characteristic velocity and $T_r$ is the reference temperature ($T_r$ can be viewed as the mean temperature of the channel), one can express (S4) in the following dimensionless form

$$A^2 \left[ Pe_T \left( \frac{\partial \theta}{\partial \bar{t}} + \bar{u}\frac{\partial \theta}{\partial X} \right) - \frac{1}{\mathcal{K}}\frac{\partial}{\partial X}\left(\mathcal{K}\frac{\partial \theta}{\partial X}\right) \right]$$
$$= \frac{1}{\mathcal{K}}\frac{\partial}{\partial Z}\left(\mathcal{K}\frac{\partial \theta}{\partial Z}\right) + A^2 \frac{\varepsilon D \Delta T \kappa^2}{\mathcal{K}}\left(\frac{E}{\partial T/\partial X}\right)^2 + \frac{\eta U^2}{\mathcal{K}\Delta T}\left(\frac{\partial \bar{u}}{\partial Z}\right)^2. \quad (S5)$$

In (S5), $Pe_T = \rho c_p U l/\mathcal{K}$ is the thermal Péclet number and $\kappa = \sqrt{2e^2 v^2 n^*/\varepsilon k_B T}$ is the Debye parameter. Unless otherwise stated, all thermophysical properties are evaluated at $T = T_r$. The induced electric field $E$, causing the Joule heating (second term on the RHS of (S5)), is obtained either from Eq. (6) or (7) (under the DH limit). Hence, the maximum of the source term describing Joule heating is obtained when $\left|E/(\partial T/\partial x)\right| \approx \left|S_{\kappa h \to 0}\right|$, since the largest thermoelectric potential is achieved under maximum confinement. To assess the significance of Joule heating for TAEs, we consider the thermophysical properties of room-temperature ionic liquids (RTILs). Typically, for RTILs at $T_r = 308$ K, $\varepsilon \approx \mathcal{O}(10) \cdot 8.854 \cdot 10^{-12}$ F/m, $D \approx \mathcal{O}(10^{-9})$ m$^2$/s, $\mathcal{K} \approx \mathcal{O}(10^{-1})$ W/(m·K) while $\kappa^{-1} \approx \mathcal{O}(10^{-9})$ m [S2–S4]. Considering $\Delta T \approx \mathcal{O}(10)$ K, $E_d \approx \mathcal{O}(10) k_B T$, $\zeta \approx \mathcal{O}(10)$ mV and $l \approx \mathcal{O}(1)$ μm, one finds $\varepsilon D \Delta T \kappa^2 \left[ E/(\partial T/\partial x) \right]^2 / \mathcal{K} \lesssim \mathcal{O}(10^{-6})$. Further multiplication of this term by a small parameter $A^2 (\ll 1)$ ensures that the influence of Joule heating on the temperature distribution is indeed negligibly small.

To determine the relative importance of advection and viscous dissipation in (S5), one needs to provide an estimate of the characteristic velocity $U$ first. Under the present flow conditions, (S2) simplifies to

$$\eta \frac{\partial^2 u}{\partial z^2} = \varepsilon E \frac{\partial^2 \psi}{\partial z^2} + \frac{1}{2} \varepsilon_T \left( \frac{\partial \psi}{\partial z} \right)^2 \frac{\partial T}{\partial x}, \tag{S6}$$

where $\varepsilon_T = d\varepsilon/dT$. (S6) reflects that the fluid motion is due to the combined influences of electroosmosis (induced by $E$, the first term on the RHS of (S6)) and thermoosmosis (generated by the temperature-dependent dielectric permittivity, the second term on the RHS of (S6)). The flow due to a temperature-dependent dielectric permittivity is known from AC electrokinetics, where it is also termed "electrothermal convection". Inserting the second derivative of $\psi$, evaluated under the DH approximation, one finds

$$\frac{\partial^2 u}{\partial z^2} = \frac{\varepsilon E \zeta \kappa^2}{\eta} \frac{\cosh(\kappa z)}{\cosh(\kappa h)} + \frac{1}{2} \frac{\varepsilon_T \zeta^2 \kappa^2}{\eta} \frac{\sinh^2(\kappa z)}{\cosh^2(\kappa h)} \frac{\partial T}{\partial x}. \tag{S7}$$

Considering the no-slip condition $\left( u \big|_{z=\pm h} = 0 \right)$ at the channel walls, double integration of (S7) with respect to $z$ yields

$$u = -\frac{\varepsilon E \zeta}{\eta} \left[ 1 - \frac{\cosh(\kappa z)}{\cosh(\kappa h)} \right] + \frac{1}{8} \frac{\varepsilon_T \zeta^2}{\eta} \frac{\partial T}{\partial x} \left[ \frac{\sinh^2(\kappa z) - \sinh^2(\kappa h) + \kappa^2 (h^2 - z^2)}{\cosh^2(\kappa h)} \right]. \tag{S8}$$

To proceed, the expression given by (7) is substituted for $E$ in (S8). Note that in (7), the Seebeck coefficient was derived under the DH approximation. Therefore, the expression for $u$ is expected to remain accurate only for small values of $\zeta$, typically for $|ev\zeta/k_B T|<1$. Correspondingly, neglecting terms $\mathcal{O}(\zeta^2)$ in (7) for such small $\zeta$, (S8) becomes

$$u = u_{HS}\left\{\left(\frac{E_d}{k_B T}-1\right)\frac{\tanh(\kappa h)}{\kappa h}\left[1-\frac{\cosh(\kappa z)}{\cosh(\kappa h)}\right]\right.$$
$$\left.+\frac{1}{8}\frac{\varepsilon_T T}{\varepsilon}\left[\frac{\sinh^2(\kappa z)-\sinh^2(\kappa h)+\kappa^2(h^2-z^2)}{\cosh^2(\kappa h)}\right]\right\}. \quad (S9)$$

In (S9), $u_{HS} = \varepsilon\zeta^2(\partial T/\partial x)/\eta T$ is the thermally induced characteristic Helmholtz-Smoluchowski velocity. The largest velocity $u_{max}$ occurs at the center plane $z=0$, and is given by

$$u_{max} = u_{HS}\left\{\left(\frac{E_d}{k_B T}-1\right)\frac{\tanh(\kappa h)}{\kappa h}\left[1-\frac{1}{\cosh(\kappa h)}\right]+\frac{1}{8}\frac{\varepsilon_T T}{\varepsilon}\left[\frac{\kappa^2 h^2}{\cosh^2(\kappa h)}-\tanh^2(\kappa h)\right]\right\}. \quad (S10)$$

Equation (S10) expresses that for a specified set of parameters $\{\varepsilon,\varepsilon_T,E_d,T\}$, $u_{max}=f(\kappa h)$. Typically, for RTILs $\varepsilon_T \approx \mathcal{O}(10^{-2})\,\text{K}^{-1}$ [S5]. Considering $T=T_r=308\,\text{K}$ along with the values of the other parameters mentioned above, the largest velocity for pure electroosmotically driven flow $u_{EOF}$ (hypothetically considering $\varepsilon_T=0$) is found to occur at $\kappa h=1.8$, yielding $u_{EOF} \approx 3.05 u_{HS}$. On the other hand, for pure thermoosmotic flow (considering $E=0$) the largest velocity occurs at $\kappa h=18$, resulting in $u_{TOF} \approx 0.03 u_{HS}$. Based on the scaling $h(\partial T/\partial x) \approx A\Delta T$ and $U \approx 3.05 u_{HS}$, the magnitude of $Pe_T$ is, therefore, limited to

$$Pe_T \lesssim 3.05\frac{\varepsilon\zeta^2}{\eta\alpha}\frac{\Delta T}{T}, \quad (S11)$$

where $\alpha(=\mathcal{K}/\rho c_p)$ is the thermal diffusivity estimated at the reference temperature $T_r$. For RTILs, $\eta \approx \mathcal{O}(10^{-1})\,\text{Pa·s}$ [S6] and $\alpha \approx \mathcal{O}(10^{-7})\,\text{m}^2/\text{s}$ [S7]. Considering $\zeta \approx \mathcal{O}(10)\,\text{mV}$ and $\Delta T \approx \mathcal{O}(10)\,\text{K}$, it is found that $Pe_T \lesssim \mathcal{O}(10^{-7})$. Further multiplication of this non-dimensional parameter by $A^2$ indicates that advection is negligible for computing the temperature distribution.

Following an identical approach, the magnitude of viscous dissipation (the third term on the RHS of (S5)) is found to be limited to

$$\frac{\eta U^2}{\mathcal{K}\Delta T} \lesssim \frac{\Delta T}{\eta \mathcal{K}} \left( \frac{3.05 \varepsilon A \zeta^2}{hT} \right)^2. \tag{S12}$$

Equation (S12) indicates that to leading order in $A$, the effects of viscous dissipation can only be neglected when

$$h \gtrsim 3.05 \frac{\varepsilon \zeta^2}{T} \sqrt{\frac{\Delta T}{\eta \mathcal{K}}}. \tag{S13}$$

Using the parameter values from above, one finds $h \gtrsim \mathcal{O}(10^{-15})$ m. Clearly, this dimension is much smaller compared to any confinement of physical relevance, thus allowing us to ignore the effects of viscous dissipation in this analysis. To leading order in $A$, the energy equation (S5), therefore, simplifies to

$$\frac{1}{\mathcal{K}} \frac{\partial}{\partial Z} \left( \mathcal{K} \frac{\partial \theta}{\partial Z} \right) = 0. \tag{S14}$$

Given the symmetry condition along the center plane, $\partial \theta / \partial Z = 0$ at $Z = 0$ and thermally insulated walls at $Z = \pm 1$, this signifies that $T = T(x)$ and $\nabla T = (\Delta T / l, 0)$.

## §2: Simplification of the Nernst-Planck Equations

To cast the NPE (S1) in non-dimensional form, we introduce the following additional dimensionless variables: $N_k = n_k^* / n_r^*$, $\overline{v} = v_k / v$, $\overline{\phi} = e v \phi / k_B T_r$ and $\overline{\psi} = e v \psi / k_B T_r$ apart from those defined earlier in the context of the derivation of (S5). Replacing the dimensional quantities in (S1) with their non-dimensional counterparts, the NPE read

$$A^2 \left[ Pe_k \left( \frac{\partial N_k}{\partial \overline{t}} + \overline{u} \frac{\partial N_k}{\partial X} \right) - \frac{1}{D_k} \frac{\partial}{\partial X} \left( D_k \frac{\partial N_k}{\partial X} + D_k \frac{\overline{v}_k N_k}{\hat{\theta}+1} \frac{\partial \overline{\phi}}{\partial X} \right) \right]$$
$$= \frac{1}{D_k} \frac{\partial}{\partial Z} \left( D_k \frac{\partial N_k}{\partial Z} + D_k \frac{\overline{v}_k N_k}{\hat{\theta}+1} \frac{\partial \overline{\phi}}{\partial Z} \right) + A^2 \left( \mathcal{G}_k \frac{l^2}{D_k} \right) \frac{1}{n_r}, \tag{S15}$$

where $Pe_k = Ul/D_k$ is the ionic Péclet number and $\hat{\theta} = \theta \Delta T / T_r$. In (S15), $\mathcal{G}_k l^2 / D_k$ can be identified as the Damköhler number (Da), the ratio between the diffusion timescale $l^2 / D_k$ and the dissociation timescale $\mathcal{G}_k^{-1}$ of the charge-neutral ion clusters. The magnitude of Da indicates whether or not the ion-cluster dissociation process attains its equilibrium quasi-instantaneously. For RTILs, the ion-cluster dissociation timescale reported in [S8] is of $\mathcal{O}(1)$ μs, while $D_k$ is usually of $\mathcal{O}(10^{-12}-10^{-9})$ m²/s [S3,S9–S11]. This gives Da $\sim \mathcal{O}(10^3-10^6) \gg 1$, suggesting that the dissociation process attains its equilibrium quasi-instantaneously. The term $\mathcal{G}_k$ is, therefore, dropped from the NPE in the further analysis.

The magnitude of $Pe_k$ can be estimated based on $U \approx 3.05 u_{\text{HS}}$. This yields

$$Pe_k \lesssim 3.05 \frac{\varepsilon \zeta^2}{\eta D_k} \frac{\Delta T}{T}. \tag{S16}$$

Considering $D_k = D \approx \mathcal{O}(10^{-12}-10^{-9})$ m²/s, along with the values of the other parameters referred to above, one finds $Pe_k \lesssim \mathcal{O}(10^{-5}-10^{-2})$. Therefore, one can safely ignore the advective terms proportional to $Pe_k$ in (S15). To leading order in $A$, further neglecting the terms of $\mathcal{O}(A^2)$, (S15) simplifies to

$$\frac{1}{D_k} \frac{\partial}{\partial Z} \left( D_k \frac{\partial N_k}{\partial Z} + D_k \frac{\overline{v}_k N_k}{\hat{\theta}+1} \frac{\partial \overline{\phi}}{\partial Z} \right) = 0. \tag{S17}$$

With $\partial \overline{\phi}/\partial Z = \partial \overline{\psi}/\partial Z$, after integration with respect to $Z$, and given the symmetry condition along the center plane at $Z = 0$, the dimensional form of (S17) is given by

$$\frac{\partial n_k^*}{\partial z} + \frac{e v_k n_k^*}{k_B T} \frac{\partial \psi}{\partial z} = 0. \tag{S18}$$

Integrating (S18) and considering the ECC concentration for $\psi = 0$ to be $n_k^* = n_b \exp(-E_d/k_B T)$, the charge carrier distribution can be expressed by

$$n_k^* = n_b \exp\left(-\frac{E_d + e v_k \psi}{k_B T}\right). \tag{S19}$$

Under chemical and thermal equilibrium, thermodynamics dictates that the concentrations of the species appearing on both sides of a chemical reaction equation are related by the Boltzmann distribution [S12]. The use of Boltzmann statistics here for the concentration of ECCs is based on the fast enough chemical equilibrium between the ECCs and the charge-neutral ion clusters, indicated by a large magnitude of Da, as discussed above. Equation (5) is derived by inserting (S19) in (S15) and neglecting the advective terms.

§3: Simplification of the Poisson Equation

In non-dimensional form, the Poisson equation (S3) can be rewritten as

$$A^2\left(\frac{\partial \bar{\varepsilon}}{\partial X}\frac{\partial \bar{\phi}}{\partial X}+\bar{\varepsilon}\frac{\partial^2 \bar{\phi}}{\partial X^2}\right)+\frac{\partial \bar{\varepsilon}}{\partial Z}\frac{\partial \bar{\phi}}{\partial Z}+\bar{\varepsilon}\frac{\partial^2 \bar{\phi}}{\partial Z^2}=-\frac{1}{2}\bar{\kappa}_r^2\sum_{k=1}^{K}\bar{\nu}N_k, \qquad (S20)$$

where $\bar{\varepsilon}=\varepsilon/\varepsilon_r$ and $\bar{\kappa}_r=\kappa_r h$, with $\varepsilon_r$ and $\kappa_r$ being the dielectric permittivity and Debye parameter at temperature $T_r$. The other non-dimensional parameters appearing in (S20) were defined earlier during the derivation of (S5) or (S15). Equation (S20) shows that to the leading order in $A$, the volumetric charge density affects the electric field only in the transverse ($z$) direction. Moreover, since $\partial \theta/\partial Z=0$ (or equivalently $\partial T/\partial z=0$), the Poisson equation in dimensional form simplifies to

$$\varepsilon\frac{\partial^2 \psi}{\partial z^2}\approx -e\sum_{k=1}^{K}\nu_k n_k^*, \qquad (S21)$$

where $\partial \phi/\partial z=\partial \psi/\partial z$ was used. Replacing $n_k^*$ in (S21) by (S19) for $k=(+,-)$ and assuming $\nu_\pm=\pm\nu$ for the oppositely charged ions, one finds

$$\frac{\partial^2 \Psi}{\partial z^2}\approx \kappa^2 \sinh(\Psi), \qquad (S22)$$

where $\Psi=\dfrac{e\nu\psi}{k_B T}$ and $\kappa$ is the local Debye parameter at temperature $T$. Within the Debye-Hückel approximation ($|e\nu\zeta/k_B T|<1$), the RHS of (S22) can be linearized, resulting in

$$\frac{\partial^2 \psi}{\partial z^2}\approx \kappa^2 \psi. \qquad (S23)$$

$\kappa$ is related to its value at the channel entrance $\kappa_0$ maintained at temperature $T_0$ by

$$\kappa=\kappa_0\sqrt{T_0 \exp(E_d \Delta T/k_B T T_0)/T}. \qquad (S24)$$

For small $\Delta T \left( \Delta T/T_0 \ll 1 \right)$, (S24) can be linearized to yield

$$\kappa = \kappa_0 + \frac{1}{2}\frac{E_d(T-T_0)}{k_B T_0^2}\kappa_0 - \frac{1}{2}\frac{(T-T_0)}{T_0}\kappa_0. \tag{S25}$$

Equation (S25) describes the deviation of local Debye parameter from its reference value $\kappa_0$, owing to thermally-driven charge carrier generation $(E_d > 0)$ and the temperature dependence of $D_k/\omega_k$.

## §4: Derivation of the Seebeck coefficient

The induced thermovoltage in form of the Seebeck coefficient $S = E/(dT/dx)$ can be computed by equating the electric current $I = \int_0^h \sum_{k=1}^K e v_k j_{k,x}\, dz$ to zero. Replacing $j_{k,x}$ with the expression from Eq. (5) for $k = (+,-)$ yields

$$\int_0^h \left[ (E_d + ev_+\psi)\frac{ev_+ n_+^* D_+}{k_B T^2} + (E_d + ev_-\psi)\frac{ev_- n_-^* D_-}{k_B T^2} \right] dz \\ - \frac{E}{dT/dx}\int_0^h \frac{e^2}{k_B T}\left( v_+^2 n_+^* D_+ + v_-^2 n_-^* D_- \right) dz = 0. \tag{S26}$$

Inserting $n_\pm^*$ expressed by Eq. (3) and assuming equal Fickian diffusion coefficients $D_\pm = D$ and $v_\pm = \pm v$, one finds

$$-\frac{E_d}{T}\int_0^h \left[ \exp\left(\frac{ev\psi}{k_B T}\right) - \exp\left(-\frac{ev\psi}{k_B T}\right) \right] dz + \frac{ev}{T}\int_0^h \psi \left[ \exp\left(\frac{ev\psi}{k_B T}\right) + \exp\left(-\frac{ev\psi}{k_B T}\right) \right] dz \\ - \frac{E}{dT/dx}\int_0^h ev \left[ \exp\left(\frac{ev\psi}{k_B T}\right) + \exp\left(-\frac{ev\psi}{k_B T}\right) \right] dz = 0. \tag{S27}$$

From that, after some further algebra, Eq. (6) is obtained. For small $\zeta$ ($|ev\zeta/k_B T| < 1$), the terms $\mathcal{O}(\zeta^n, n \geq 2)$ can be ignored in (7), yielding

$$S_{DH} = -\frac{E_d \zeta}{k_B T^2}\frac{\tanh(\kappa h)}{\kappa h} + \frac{\zeta}{T}\frac{\tanh(\kappa h)}{\kappa h}. \tag{S28}$$

For small $\Delta T$ ($\Delta T/T_0 \ll 1$), the Seebeck coefficient (S28) can be further simplified to obtain Eq. (8).

**§5: Effect of Stern layer conductivity on the Seebeck coefficient**

It was reported that at the interface between an ionic liquid and a solid surface, a Stern layer with a characteristic thickness $\delta$ of a few ion diameters will form [S13]. The ions within this layer remain strongly attached to the surface. Above this Stern layer there is a diffuse layer of ECCs. The Seebeck coefficient computed so far is based on the flux within this diffuse layer, thus ignoring the influence of the Stern layer. However, the ions in the Stern layer are associated with a surface conductance $\sigma_s$ and, therefore, can give rise to an electric current. Incorporating the contributions of both bulk (diffuse layer) and surface (Stern layer) conduction, the total electric current $\mathcal{I}$ can be expressed by

$$\mathcal{I} = \int_0^{h'} \sum_{k=1}^{K} e v_k j_{k,x} dz + \int_{h'}^{h} \sigma_s E dz, \quad (S29)$$

where $h - h' = \delta$. Replacing $j_{k,x}$ in (S29) with (5), (S29) becomes

$$\mathcal{I} = \int_0^{h'} \sum_{k=1}^{K} \left[ -e v_k n_k^* D_k \left( \frac{E_d + e v_k \psi}{k_B T^2} \right) \frac{dT}{dx} + \frac{e^2 v_k^2 n_k^* D_k}{k_B T} E \right] dz + \int_{h'}^{h} \sigma_s E dz. \quad (S30)$$

Setting $\mathcal{I} = 0$ results in

$$\frac{E}{dT/dx} = \frac{\dfrac{e}{k_B T^2} \int_0^{h'} \sum_{k=1}^{K} v_k n_k^* D_k (E_d + e v_k \psi) dz}{\dfrac{e^2}{k_B T} \int_0^{h'} \sum_{k=1}^{K} n_k^* D_k v_k^2 dz + \int_{h'}^{h} \sigma_s dz}. \quad (S31)$$

Assuming $D_k = D$ and $v_k = \pm v$, the integrals in (S31) can be evaluated under the DH approximation, yielding

$$S = \frac{-\dfrac{E_d \zeta}{k_B T^2} \dfrac{\tanh(\kappa h)}{\kappa h} + \dfrac{\zeta}{T} \dfrac{\tanh(\kappa h)}{\kappa h} \left[ 1 + \dfrac{1}{2}\left(\dfrac{e v \zeta}{k_B T}\right)^2 \left\{ \dfrac{\tanh^2(\kappa h)}{3} + \dfrac{1}{\cosh^2(\kappa h)} \right\} \right]}{1 + \left(\dfrac{e v \zeta}{2 k_B T}\right)^2 \left\{ \dfrac{\tanh(\kappa h)}{\kappa h} + \dfrac{1}{\cosh^2(\kappa h)} \right\} + \dfrac{\sigma_s \delta}{\varepsilon \kappa^2 h D}}, \quad (S32)$$

where the different symbol for the Seebeck coefficient indicates the fact that the Stern layer contribution was taken into account. Equation (S32) demonstrates that $\sigma_s$ reduces the Seebeck coefficient, and hence the thermoelectric potential. By how much the Seebeck coefficient will be reduced remains an open question unless a model for computing the surface conductance is available. This is a challenging task that probably requires approaches going beyond a continuum-mechanical description. We would also like to point out that there may be an additional effect to consider that, by contrast to what (S32) predicts, yields an increase of the thermovoltage. Ions in the Stern layer may be transported via thermally-induced hopping in a potential energy landscape created by the solid surface, as suggested in [S14]. In conclusion, the effects due to the Stern layer remain an open issue that deserves closer attention in the future.

**§6: Estimation of the surface charge density based on the zeta potential**

The electrostatic potential can be analytically determined by solving the non-linear Poisson-Boltzmann equation (S22) [S15]. At the center plane of the slit channel, this is given by

$$\psi\big|_{z=0} = \zeta - 2\frac{k_B T}{ev} \ln\left[\mathrm{cd}\left(\mathcal{L}\big|\mathfrak{D}\right)\right] \tag{S33}$$

with

$$\mathcal{L} = \frac{\kappa h}{2} \exp\left(-\frac{ev}{2k_B T}\psi\big|_{z=0}\right) \tag{S34}$$

and

$$\mathfrak{D} = \exp\left(\frac{2ev}{k_B T}\psi\big|_{z=0}\right). \tag{S35}$$

In (S33), cd stands for a Jacobi elliptic function with argument $\mathcal{L}$ and parameter $\mathfrak{D}$. Based on that, the surface charge density $q$ is given by [S16]

$$q = -\frac{\varepsilon\kappa k_B T}{ev}\left(\mathfrak{D}^{3/4} - \mathfrak{D}^{-1/4}\right)\frac{\mathrm{sn}\left(\mathcal{L}\big|\mathfrak{D}\right)}{\mathrm{cn}\left(\mathcal{L}\big|\mathfrak{D}\right)\mathrm{dn}\left(\mathcal{L}\big|\mathfrak{D}\right)}, \tag{S36}$$

where $\mathrm{sn}\left(\mathcal{L}\big|\mathfrak{D}\right)$, $\mathrm{cn}\left(\mathcal{L}\big|\mathfrak{D}\right)$ and $\mathrm{dn} = \mathrm{cn}/\mathrm{cd}$ again denote Jacobi elliptic functions of argument $\mathcal{L}$ and parameter $\mathfrak{D}$. $\kappa$ in (S34) and (S36) is expressed by (S24).

For a given parameter set $\{\varepsilon, \kappa, T\}$, Eqs. (S33)-(S36) establish an implicit relation between $q$ and $\zeta$. To compute $q$, first $\psi|_{z=0}$ is determined by simultaneously solving (S33)–(S36). Inserting this $\psi|_{z=0}$ in (S36), one can evaluate $q$. The Jacobi functions are computed by employing the ELLIPJ function of Matlab (Version 9.10.0.1602886, R2021a). Finally, the surface charge density determined in this way is used as a Neumann boundary condition for the numerical simulation of the PNP equations to compute $S$.

### §7: Numerical scheme for the solution of Eqs. (6)

The DH approximation limits the accuracy of the analytical solution of (S22) and hence, the corresponding Seebeck coefficient only to small values of $\zeta$ ($|ev\zeta/k_B T| < 1$). A numerical solution is therefore essential to compute reliable values of $S$ for large zeta potentials. Using the BVP4C function in Matlab, Eq. (S22) is solved by employing the boundary conditions $\partial \psi / \partial z = 0$ (at $z = 0$) and $\psi = \zeta$ (at $z = \pm h$). Based on a grid independence study, a grid with 1000 points is selected in the $z$-direction. $\psi$ computed in this manner is inserted in (6), and a numerical integration is carried out using the Simpson's 1/3 rule for a wide range of $\kappa_0 h$. $S$ evaluated following this numerical approach is found to be practically indistinguishable from the analytical solution (within its domain of validity i.e., for $|ev\zeta/k_B T| < 1$).

### §8: Numerical simulation framework for the Poisson-Nernst-Planck (PNP) equations

The Seebeck coefficient for TAEs is numerically computed by solving the governing equations in two dimensions. The computational domain is a slender rectangle with a reservoir at each end, as shown in Fig. 1a. The coupled PNP equations (Eqs. (1) and (2) in the main text) are solved in conjunction with the energy equation $\nabla^2 T = 0$ (ignoring the terms with negligible contributions, as discussed in §1). Given the mirror symmetry along the $z = 0$ plane, only the upper half of the domain is considered. The length $l$ is taken as $250h$, while the reservoir dimensions (length×height) are $20h \times 10h$. The computations are carried out with the steady-state PDE mode of Comsol Multiphysics 5.6.0, employing the PARDISO solver. A structured mesh with rectangular elements and quadratic Lagrangian shape functions is used, employing refinement close to corners and boundaries. A grid independence check was carried out, and the results are found to be virtually invariant under further mesh refinements at $\sim 6 \times 10^4$ mesh elements. The iterative solution process is terminated at a relative tolerance (i.e. the maximum allowable relative error in the solution, as reported by the Comsol solver) level of $10^{-3}$.

The computations are carried out by fixing the temperature of the left wall of the leftmost reservoir at $T_0 = 298\,\text{K}$, and maintaining $\Delta T = 25\,\text{K}$ between the extreme end walls of both reservoirs. The walls of the channel and the reservoirs are considered to be thermally insulated and impermeable for the solute and the dissolved ions. Only the channel walls carry a net electric charge. The potential $\phi$ of the rightmost wall of the right reservoir is set to zero and varied for the leftmost wall of the left reservoir, so that the net electric current $I$ vanishes. The particular value of $\phi$ for which $I$ becomes zero is the thermoelectric potential for a given set of the parameters $\kappa_0$, $E_d$, as well as $\zeta$ or $q$.

**§9: Effect of a temperature-dependent dielectric permittivity on the Seebeck coefficient**

The dielectric permittivity $\varepsilon$ of a medium usually varies with temperature. Literature suggests that for most fluids, $\varepsilon$ decreases with increasing temperature. This variation of $\varepsilon$ affects the electrostatic potential distribution within the confinement via the Poisson equation (Eq. (2) in the main text). Therefore, a corresponding change in the Seebeck coefficient can also be expected.

While the variation of dielectric permittivity with temperature is well documented in the case of dilute aqueous electrolytes, data is scarce for TAEs. Some studies [S5] report a linear decrement in $\varepsilon$ with increasing temperature over a comparatively large temperature interval, whereas others [S17] suggest a sharp change in $\varepsilon$ within a narrow temperature interval. No general statement can, therefore, be made regarding the variation of dielectric permittivity in TAEs. To assess the influence of dielectric permittivity variations on the Seebeck coefficient, we here consider an extreme situation, similar to that reported in [S17], where $\varepsilon$ jumps from 21 to 7.5 within a temperature interval of about 1K.

The results of the corresponding numerical computations, shown in Fig. S.1, reveals that, even under such extreme conditions, the Seebeck coefficient does not vary appreciably compared to the case of a fixed $\varepsilon$. Therefore, the Seebeck coefficient we have computed is expected to give a reasonably accurate picture of the thermoelectric response even of TAEs in which the dielectric permittivity exhibits a strong temperature dependence.

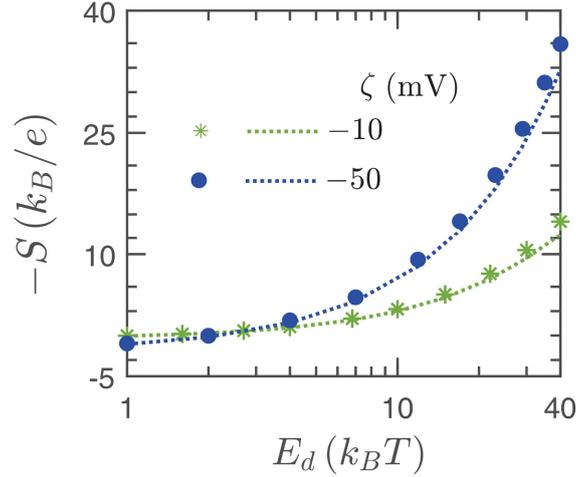

FIG. S1. Effect of a temperature-dependent dielectric permittivity on the Seebeck coefficient. $S$ is plotted as a function of the activation energy for four different cases: fixed $\varepsilon$, and a temperature dependent $\varepsilon$, each evaluated at two different values of the wall $\zeta$ potential. Symbols indicate the data computed considering a linear permittivity variation between $\varepsilon = 21$ at the cold reservoir and $\varepsilon = 7.5$ at the hot reservoir for a temperature drop of $\Delta T = 1\,\text{K}$ between both reservoirs. For each $\zeta$ potential, the dotted line represents $S$ computed for fixed $\varepsilon\,(=14.25)$, the average between 21 and 7.5. All data have been obtained for $\kappa_0 h = 0.1$ and $T_0 = 298\,\text{K}$.